\documentstyle{mn}

\newcounter{parentequation}
\def\beglet{
  \addtocounter{equation}{1}%
  \setcounter{parentequation}{\value{equation}}%
  \setcounter{equation}{0}%
  \def\theequation{\arabic{parentequation}\alph{equation}}%
  \ignorespaces
}
\def\endlet{
  \setcounter{equation}{\value{parentequation}}%
  \def\theequation{\arabic{equation}}%
}

\def\ltsima{$\; \buildrel < \over \sim \;$}
\def\gtsima{$\; \buildrel > \over \sim \;$}
\def\simlt{\lower.5ex\hbox{\ltsima}}
\def\simgt{\lower.5ex\hbox{\gtsima}}

\def\hmpc{h^{-1} {\rm Mpc}}
\def\hmpci{h\; {\rm Mpc}^{-1}}
\def\etal{{\rm et al.}~\rm}

\begin{document}

\title[CMB and the 2dF Galaxy Redshift Survey]
{Evidence for a non-zero $\Lambda$ and a low matter density from a 
combined analysis of  the 2dF Galaxy Redshift Survey
and Cosmic Microwave Background Anisotropies}

\author[G. Efstathiou  et al.]{ G. Efstathiou$^{1,2}$, Stephen Moody$^{1}$,
John A. Peacock$^{3}$, 
Will J. Percival$^{3}$, \cr
Carlton Baugh$^{4}$,  
Joss Bland-Hawthorn$^{5}$, 
Terry Bridges$^{5}$,  
Russell Cannon$^{5}$, \cr Shaun Cole$^{4}$, \
Matthew Colless$^{6}$,  Chris Collins$^{7}$, Warrick Couch$^{8}$, 
Gavin Dalton$^{9}$,  \cr Roberto De Propis$^{8}$, 
Simon P. Driver$^{10}$, 
Richard S. Ellis$^{11}$, Carlos S. Frenk$^{4}$, \cr Karl
Glazebrook$^{12}$,  Carole Jackson$^{6}$, Ofer Lahav$^{1}$, 
 Ian Lewis$^{5}$,
Stuart Lumsden$^{13}$, \cr Steve Maddox$^{14}$, 
Peder Norberg$^{4}$,  Bruce A. Peterson$^{6}$,  Will Sutherland$^{3}$, \cr
Keith Taylor$^{11}$ (The 2dFGRS Team)\smallskip \\
1. Institute of Astronomy, Madingley Road, Cambridge CB3 OHA. UK. \\
2. Theoretical Astrophysics, Caltech, Pasadena, CA 91125, USA. \\
3. Institute for Astronomy, University of Edinburgh, Royal Observatory, Blackford Hill, Edinburgh, EH9 3HJ, UK. \\
4. Department of Physics, University of Durham, South Road, Durham DH1 3LE, 
UK. \\
5. Anglo-Australian Observatory, P.O. Box 296, Epping, NSW 2121, Australia.\\
6. Research School of Astronomy and Astrophysics, The Australian National University, Weston Creek, ACT 2611, Australia \\
7. Astrophysics Research Institute, Liverpool John Moores University, Twelve Quays House, Birkenhead, L14 1LD, UK \\
8. Department of Astrophysics, University of New South Wales, Sydney, NSW 2052,
Australia.\\
9. Astrophysics, Nuclear and Astrophysics Laboratory, University of Oxford, Keble Road, Oxford OX1 3RH, UK. \\ 
10. School of Physics and Astronomy, University of St Andrews, North Haugh, St Andrews, Fife, KY6 9SS, UK. \\
11. Department of Astronomy, Caltech, Pasadena, CA 91125, USA. \\
12. Department of Physics and Astronomy, Johns Hopkins University, Baltimore,
MD 21218-2686, USA.\\
13. Department of Physics, University of Leeds, Woodhouse Lane, Leeds, LS2 9JT, UK. \\
14. School of Physics and Astronomy, University of Nottingham, Nottingham, NG7 2RD, UK. \\
 }

\maketitle

\vskip -0.3 truein

\begin{abstract}
We perform a joint likelihood analysis of the power spectra of the 2dF
Galaxy Redshift Survey (2dFGRS) and the cosmic microwave background
(CMB) anisotropies under the assumptions that the initial fluctuations
were adiabatic, Gaussian and well described by power laws with scalar
and tensor indices of $n_s$ and $n_t$. On its own, the 2dFGRS  sets
tight limits on the parameter combination $\Omega_m h$\footnotemark,
but relatively weak limits on the fraction of the cosmic matter
density in baryons $\Omega_b/\Omega_m$.  The CMB anisotropy data alone
set poor constraints on the cosmological constant and Hubble constant
because of a `geometrical degeneracy' among parameters. Furthermore,
if tensor modes are allowed, the CMB data allow a wide range of
values for the physical densities in baryons and cold dark matter
($\omega_b = \Omega_b h^2$ and $\omega_c = \Omega_c h^2$).  Combining
the CMB and 2dFGRS data sets helps to break both the geometrical and
tensor mode degeneracies.  The values of the parameters derived here are
consistent with the predictions of the simplest models of inflation,
with the baryon density derived from primordial nucleosynthesis
and with  direct measurements of the Hubble parameter.  In particular,
we find strong evidence for a positive cosmological constant with a
$\pm 2 \sigma$ range of $0.65 < \Omega_\Lambda < 0.85$, completely
independently of  constraints on $\Omega_\Lambda$ 
derived from Type Ia supernovae. 

\smallskip

\noindent
{\bf Key words:} Galaxy clustering,  large-scale structure, 
cosmic microwave background- cosmology: miscellaneous.

\end{abstract}

\footnotetext{{Here $h$ is Hubble's constant $H_0$ in units of
$100{\rm km}{\rm s}^{-1} {\rm Mpc}^{-1}$. The cosmic densities
in baryons, cold dark matter and vacuum energy are denoted by
$\Omega_b$, $\Omega_c$ and $\Omega_\Lambda$. The total
matter density is $\Omega_m = \Omega_b + \Omega_c$ and the
curvature is fixed by $\Omega_k = 1 - \Omega_m - \Omega_\Lambda$.}}

\section{Introduction}

Until recently, cosmology was a subject starved of data, with poor or
non-existent constraints on fundamental quantities such as the
curvature of the Universe, the power spectrum of density
irregularities.  and the cosmic densities in baryons, cold dark matter
and vacuum energy. The situation has changed dramatically over the
last few years. Following the discovery of the CMB anisotropies (Smoot
\etal 1992) it was realized that many of the fundamental parameters of
our Universe could be determined via accurate, high resolution
measurments of the CMB ({\it e.g.} Bond \etal 1994, Jungman \etal
1996).  This has now become a reality through a number of
exquisite ground based and balloon experiments (see Halverson \etal
2001; Lee \etal 2001; Netterfield \etal 2001). Constraints on
cosmological parameters derived from these experiments are described
in several recent papers (de Bernadis \etal 2001; Pryke \etal 2001;
Stompor \etal 2001; Wang, Tegmark \& Zaldarriaga 2001).

Significant advances have also been made in surveying large scale
structure in the Universe. The
 development of wide-field correctors and multi-fibre spectroscopy
means that it is now possible to measure redshifts of hundreds of
thousands of galaxies.  Two such redshift surveys are underway. The
2dF Galaxy Redshift Survey (2dFGRS) utilises the 2dF instrument at the
Anglo-Australian Telescope and is based on  a revised version of the
APM Galaxy Survey (Maddox \etal 1990)
limited at $b_J = 19.45$. Redshifts have  now been measured for over $175\
000$ galaxies (see Colless \etal 2001, for a description of this
survey). The Sloan Digital Sky Survey (SDSS, York \etal 2000) is a CCD
imaging and spectroscopic survey that aims to measure redshifts for a
sample of $900 \ 000$ galaxies. An analysis of the galaxy
power spectrum from the 2dFGRS is described by Percival \etal (2001,
hereafter P01). First results on galaxy clustering from a  subsample
of the SDSS are presented by Zehavi \etal (2001).

In addition, a number of other investigations have greatly improved
the accuracy of various cosmological parameters. For example, surveys
of high redshift Type Ia supernovae have revealed tantalizing evidence
for an accelerating Universe (Perlmutter \etal 1999, Riess \etal
1999); the HST Hubble key project has concluded that $H_0 = 72 \pm
8 \;{\rm km}\;{\rm s}^{-1} {\rm Mpc}^{-1}$ (Freedman \etal 2001); primordial
nucleosythesis and deuterium abundance measurements from quasar
absorption lines imply a baryon density $\omega_b = 0.020 \pm 0.002$
(Burles \& Tytler 1998ab; Burles, Nollett \& Turner 2001). With
these and many other ambitious projects at various stages of
development ({\it e.g.} weak shear lensing surveys, CMB
interferometers, CMB polarization experiments, the MAP, Planck and
SNAP satellites\footnote{Descriptions of these satellites can be found
on the following web pages: http://snap.lbl.gov/, http://map.gsfc.nasa.gov,
http://astro.estec.esa.nl/SA-general/Projects/Planck.}) it is clear 
that the era of quantitative
cosmology has arrived.

In this paper, we perform a combined likelihood analysis of the CMB
anisotropy data and of the 2dFGRS galaxy power spectrum measured by
P01. We assume that the initial fluctuations were Gaussian, adiabatic
and described by power-law fluctuation spectra. Matter is assumed to 
consist of baryons and cold dark matter
(CDM) and neutrinos are assumed to have negligible rest masses.
We allow tensor and
scalar modes and place no constraints on their respective spectral
indices and relative amplitudes. Almost all previous analyses of the
CMB anisotropies have neglected tensor modes. However, including
tensor modes introduces a major new degeneracy (referred to as the {\it
tensor degeneracy} in this paper) that significantly widens the range
of allowed parameters (see Efstathiou \& Bond 1999, Wang \etal 2001,
Efstathiou 2001). The tensor degeneracy can be broken by invoking
additional data sets.  Wang \etal 2001 combine the CMB data with
measurements of the galaxy power spectrum from the IRAS PSCz survey
(Hamilton, Tegmark \& Padmanabahn 2000), estimates of the power
spectrum on small scales from observations of the Ly$\alpha$ forest
(Croft \etal 2001) and limits on the Hubble constant from HST Hubble
Key Project. Here we investigate how the major parameter degeneracies
can be broken by combining the CMB data with  the 2dFGRS
power spectrum. The 2dFGRS power spectrum is based on a large survey,
with well controlled errors, and as demonstrated by P01 already sets
interesting limits on the matter content of the Universe. Our expectation
(see Efstathiou 2001) is that a joint analysis of the CMB and 2dFGRS
will produce accurate estimates of the baryonic and matter densities
of the Universe and set useful limits on a cosmological constant. This
expectation is borne out by the results described in the rest of this
paper.

\section{Likelihood analysis}

\subsection{Analysis of the 2dFGRS power spectrum}

We use the estimates of the galaxy power spectrum and associated
covariance matrix computed by P01.  As in P01, we fit these estimates
to theoretical models of the linear matter power spectrum of CDM
models using the fitting formulae of Eisenstein \& Hu (1998). The
fits are restricted to the wavenumber range $0.02  < k/(h\;{\rm Mpc}^{-1})
< 0.15$. Redshift-space distortions (see Peacock
\etal 2001) and non-linear evolution of the power spectrum have
negligible effect on the shape of the power spectrum at these
wavenumbers. We will assume that the galaxy power spectrum within this
wavenumber range is directly proportional to the linear matter power
spectrum. This is a key assumption in the analysis presented in this
paper.  The lower wavenumber limit is imposed (conservatively) to
reduce the sensitivity of the analysis to fits to the redshift
distribution of galaxies, which are computed independently for
different zones of the survey. Since the 2dFGRS has a complex
geometry, the theoretical power spectra must be convolved with the
spherical average over wavenumber of the survey `window function'.
These convolved theoretical estimates are used together with the
spherically averaged estimates of the power spectrum of the data and
the covariance matrix (computed from Gaussian realizations of the
2dFGRS) to form a likelihood function. We refer the reader to P01 for
a full discussion of each of these steps in the analysis.

In general, the linear power spectrum with wavenumber
measured in inverse {\rm Mpc} depends on the baryonic and CDM physical
densities ($\omega_b$ and $\omega_c$), the scalar spectral index $n_s$
and an overall amplitude $A$ (the amplitude is treated as an
`ignorable' parameter in this paper and so its precise definition is
unimportant). However since we use redshift to measure distances, the
wavenumber of the observations scales as $h \;{\rm Mpc}^{-1}$. The
comparison of theory with observations therefore requires the
introduction of the parameter $h$. In fact, the set of variables $A$,
$n_s$, $\Omega_m h$, $\omega_b/\omega_m$ and $h$ are natural variables
for an analysis of large-scale structure: the combination $\Omega_m h$
defines the overall shape of the CDM transfer function ( and for
negligible baryon density is sometimes denoted by the shape parameter
$\Gamma$), while the ratio $\omega_b/\omega_m$ determines the
amplitude of baryonic oscillatory features in the transfer function
(Eisenstein \& Hu 1998; Meiksin, Peacock \& White 1999).

\begin{figure}

\vskip 3.6 truein

\includegraphics{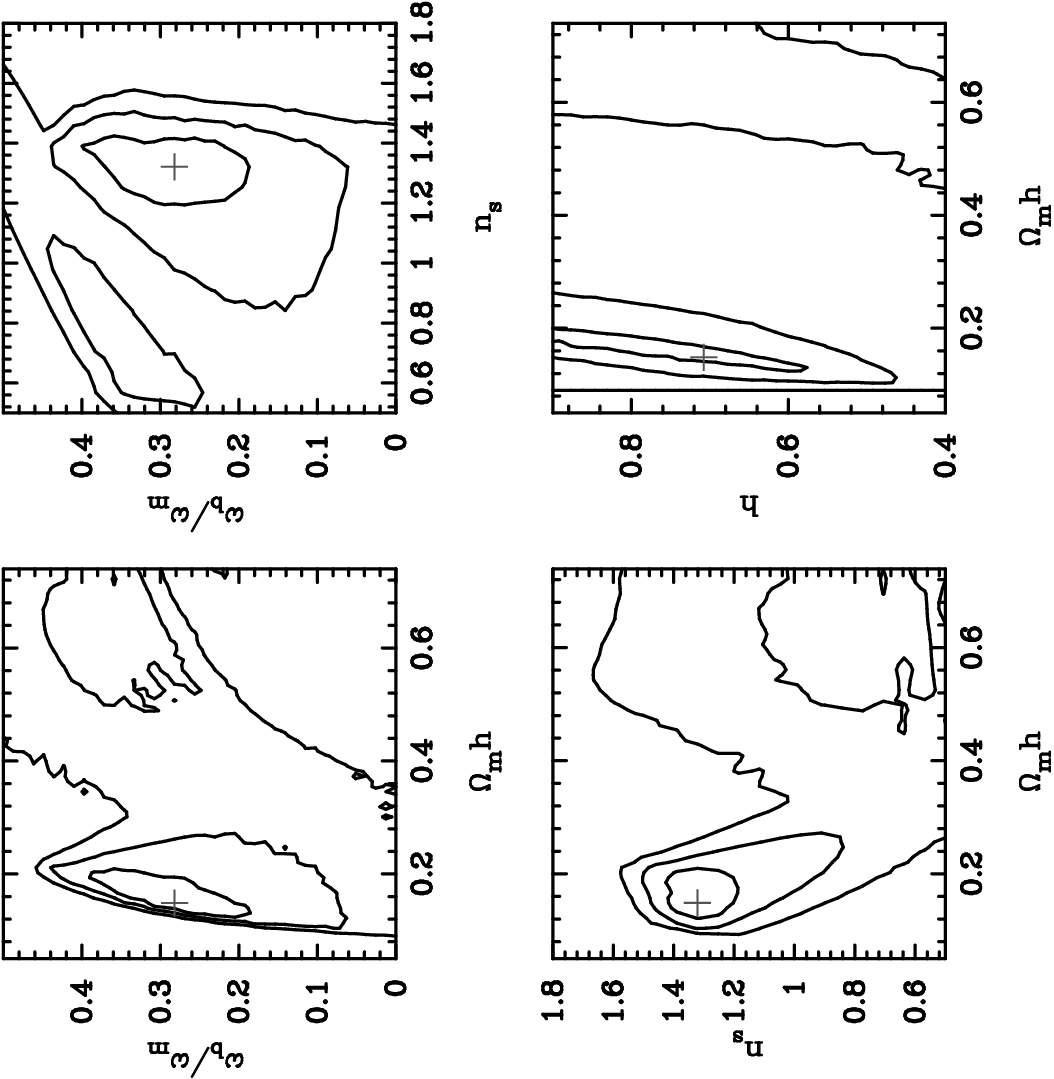}

\caption
{Contours ($1$, $2$ and $3 \sigma$) of the pseudo-marginalized likelihood
functions (see text for details) for various pairs of parameters
computed by fitting to the galaxy power spectrum of the
2dFGRS. These contours correspond to changes in the likelihood
of $2 \Delta {\rm ln} ({\cal L})$ of
$2.3$, $6.0$ and $9.2$.
The crosses show the position of  maximum likelihood. }
\label{figure1}
\end{figure}

Fig. 1 shows various two-dimensional projections of the
`pseudo-marginalized' 2dFGRS likelihood function. When using a large
number of parameters (as in the CMB and CMB+2dFGRS analyses described in
the next two subsections), it is impractical to compute marginalized
likelihood contours by numerically integrating over the likelihood
distribution. Instead, a `pseudo-marginalized' likelihood function in
$p$ out of $M$ parameters is computed by setting the remaining $M-p$
parameters at the values which maximise the likelihood. For a
multivariate Gaussian distribution, this is equivalent to integrating
over the $M-p$ parameters assuming uniform prior distributions (see
Tegmark, Zaldarriaga \& Hamilton 2001). However, the actual
likelihood distributions are not exactly Gaussian (as is evident from
the asymmetrical contours in Figs 1 and 3) and so confidence
limits assigned to pseudo-marginalized distributions are  approximate.
The contours in the
($\omega_b/\omega_m$, $\Omega_m h$) plane can be compared with Fig. 5 of
P01 where the spectral index was assumed to be scale
invariant. Relaxing the constraint on the spectral index clearly
widens the allowed range of $\omega_b/\omega_m$, but the
data still place a tight
constraint on the `shape' parameter $\Omega_m h$. 
As we will see below, the constraints on $\Omega_m h$ and
$n_s$ prove particularly important
in breaking degeneracies among parameters inherent in the
analysis of  CMB data. 

\subsection{Analysis of the CMB anisotropies}

The likelihood analysis presented here uses the compilation of band
power estimates $\Delta T_B^2$ and their covariance matrix
$C_{BB^\prime}$ (including a a model for calibration and beam errors)
computed by Wang \etal (2001) from $105$ CMB anisotropy measurements.
Each band power estimate is related to the power spectrum $C_\ell$ of
the CMB anisotropies by
\begin{equation}
 \Delta T_B^2 = {T_0^2 \over 2 \pi} \; \sum_\ell \ell(\ell+1) C_\ell
W_B(\ell) \label{2.1}
\end{equation}
where $W_B$ is the window function for each band power computed
by Wang \etal These band-power estimates are plotted in Fig. 2.

\begin{figure}

\vskip 2.9 truein

\includegraphics{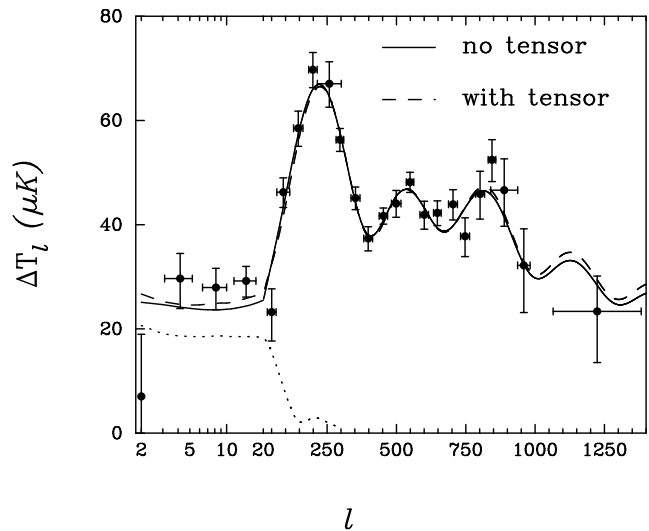}

\caption
{The points show band-averaged observational estimates of the CMB
power spectrum from Wang \etal (2001) together with $\pm 1 \sigma$
errors. The lines shows the CMB power spectra for the adiabatic
fiducial inflationary models that provide the best fit to the CMB and
2dFGRS power spectra. The parameters of these model are listed in
Table 1. The solid line shows the best fit  without a
tensor component  (fit B). The dashed line shows the best fit  (fit C)
including a tensor component (shown by the dotted line).}
\label{figure2}
\end{figure}

\begin{figure*}

\vskip 8.2  truein

\includegraphics{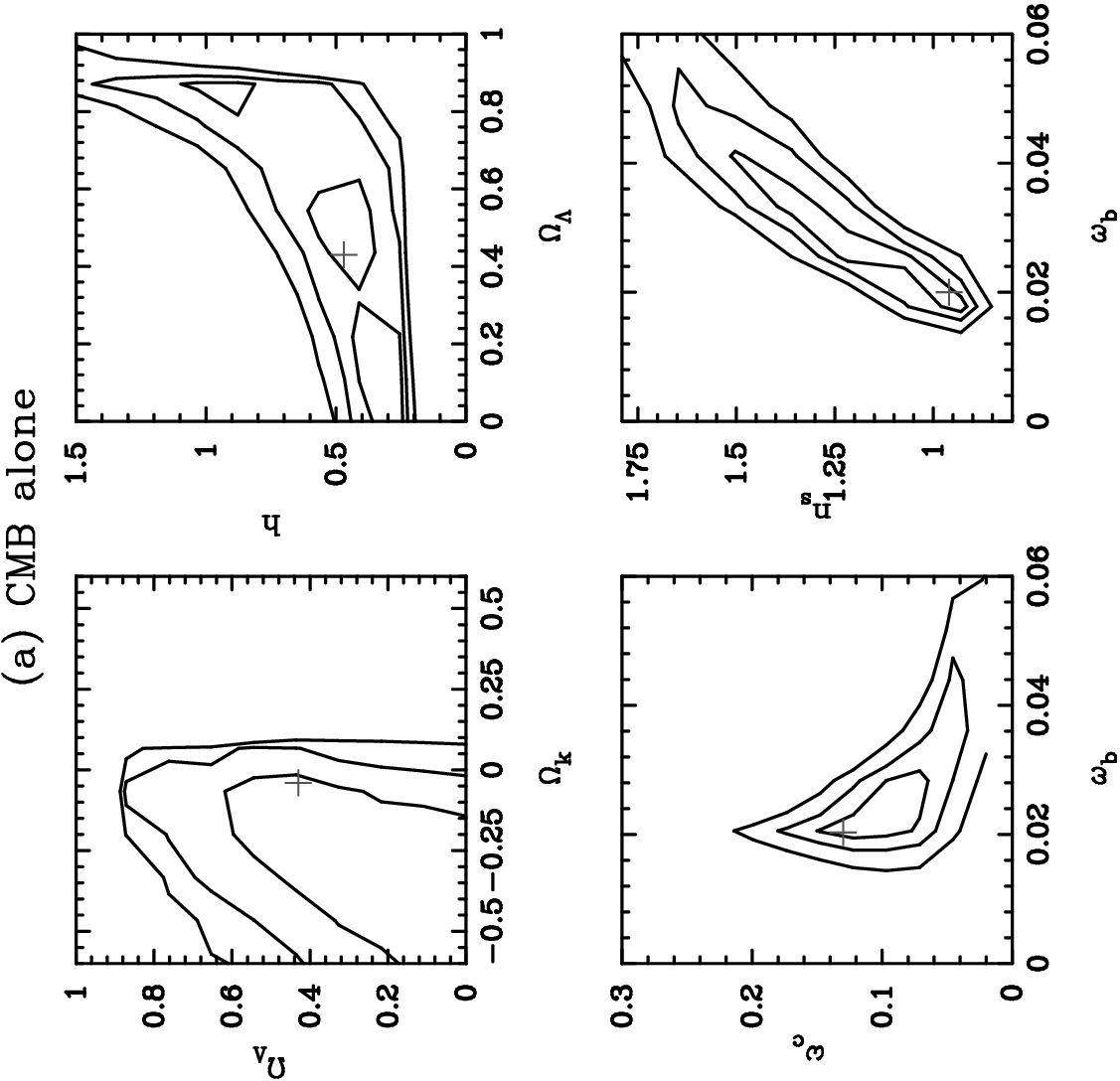}
\includegraphics{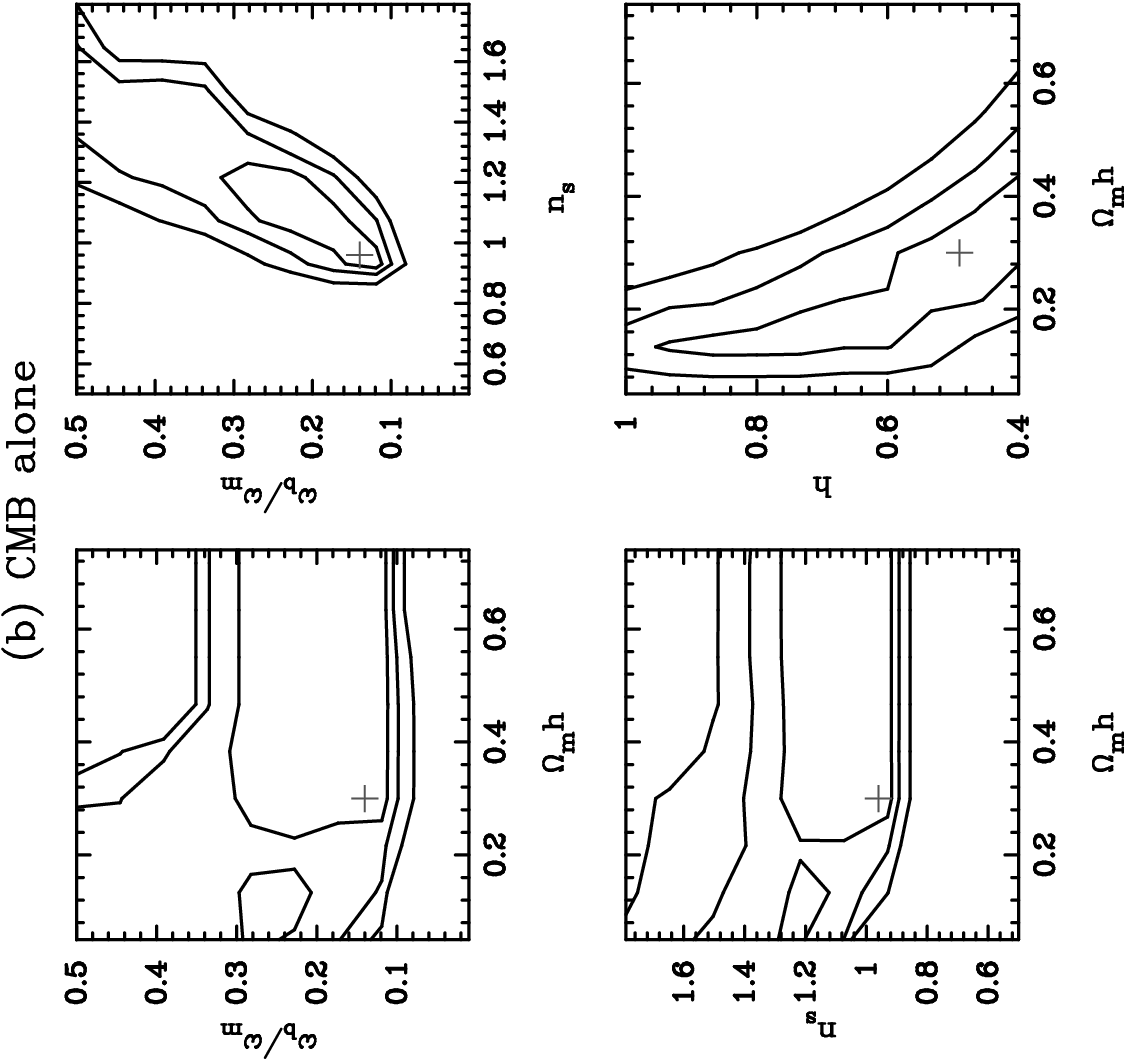}
\includegraphics{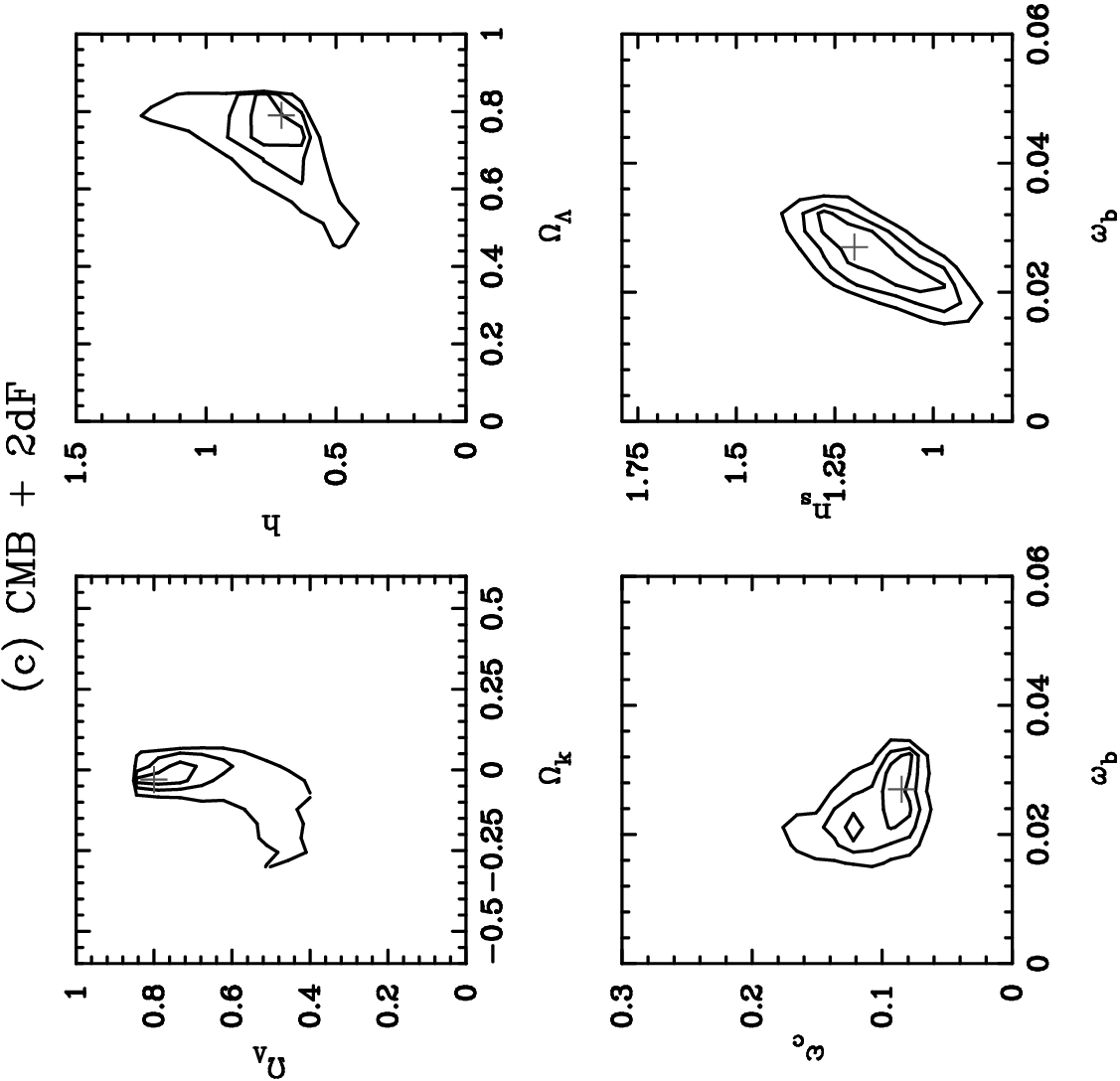}
\includegraphics{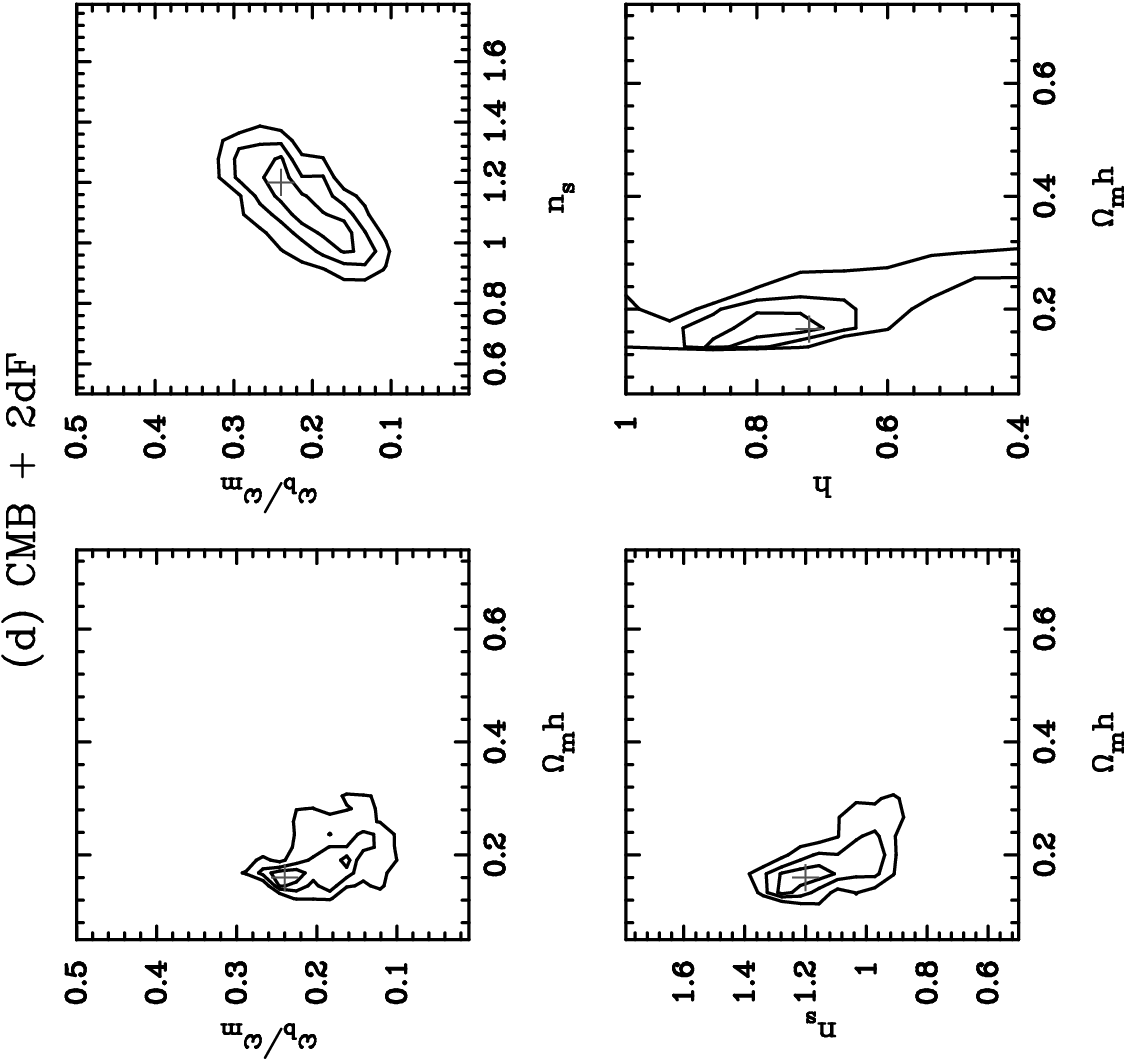}

\caption
{ Results of the nine parameter likelihood analysis.
Figs 3a and 3b show approximate $1$, $2$ and $3\sigma$ likelihood
contours for various parameter pair combinations computed from an
analysis of the CMB data alone. Figs 3a use variables natural to
the CMB analysis and illustrate the geometrical and tensor
degeneracies.  Figs 3b use the variables natural to the analysis of
the galaxy power spectrum (as used in Fig. 1). Figs 3c and 3d
show the likelihood contours of CMB and 2dFGRS data combined. The crosses
in each panel show the position of the maximum likelihood.}

\label{figure3}
\end{figure*}

The likelihood analysis of the CMB data uses nine parameters. These are:
$\omega_b$ and $\omega_c$; $\Omega_\Lambda$ and $\Omega_k$;
 the scalar and tensor spectral indices
$n_s$ and $n_t$; the optical depth to Thomson scattering $\tau_{opt}$,
assuming that the inter-galactic medium was abruptly reionized some
time after recombination; the amplitude
$Q^2$ of the scalar component and the ratio of $\overline r$ of the
tensor to scalar amplitudes. Note that definitions of the scalar
and tensor amplitudes  differ  from paper
to paper. Here we scale the scalar and tensor spectra so that
\beglet
\begin{eqnarray}
{1 \over 4 \pi}  \sum_{\ell=2}^{1000} (2\ell+1) \hat C^S_\ell & =&  (4\times10^{-5})^2, \\
{1 \over 4 \pi}   \sum_{\ell=2}^{50}  (2\ell+1) \hat C^T_\ell & =&  (2\times 10^{-5})^2, 
\end{eqnarray}
\endlet and fit to the data by scaling with the parameters $Q$ and
$\overline r$, $C_\ell = C_\ell^S + C_\ell^T = Q^2( \hat C_\ell^S + \overline r
\hat C_\ell^T)$. The numbers in equation (2) were chosen so that
models with $Q$ of approximately unity match the data points plotted
in Fig. 2 and models with $\overline r \approx 1$ have scalar and
tensor modes of comparable amplitude. We normalize the spectra in this
way to reduce the sensitivity of the normalization parameters to other
parameters that affect the low order multipole moments ({\it e.g.}
$\Omega_\Lambda$ and $\Omega_k$) and to decouple $Q$ from the optical
depth parameter $\tau_{opt}$. This method of normalizing helps to
stabilize searches for global maxima of the likelihood
functions. For our best fit models of Table 1 below we list values of
the more commonly used parameter $r_{10} \equiv C^T_{10}/C^S_{10}$ in
addition to $\overline r$.  In simple models of inflation, the
parameters $r_{10}$ (or $\overline r$), $n_s$ and $n_t$ are related to
each other (see {\it e.g.}  Hoffman \& Turner 2001 for a recent
discussion). The relations are model dependent, however, 
and can be violated in multi-field inflation models and in
superstring  inspired models such as the pre-big bang 
(Buonanno, Damour \& Veneziano 1999) and  ekpyrotic
scenarios  (Khoury, Ovrut, Steinhardt \& Turok 2001). We
therefore assume no relations between $r_{10}$, $n_s$ and $n_t$ 
in this paper.

Results from the likehood analysis of the CMB data are illustrated in Fig.
3.  Almost all of the variance of the parameters used in this
analysis, with the exception of
$Q$, comes from two major degeneracies (see Efstathiou 2001 for a
detailed discussion). These two degeneracies are illustrated by the
likelihood contours plotted in Fig. 3a. The top two panels
illustrate the `geometrical' degeneracy. This degeneracy arises
because models with identical matter content, primordial power spectra
and angular diameter distance to the last scattering surface produce
almost identical CMB power spectra. This leads to a strong degeneracy
between $\Omega_\Lambda$ and $\Omega_k$, which is broken only for
extreme values of $\Omega_\Lambda$ by the 
integrated Sachs-Wolfe effect which modifies the shape of the CMB
power spectrum at low multipoles 
(see Efstathiou \& Bond 1999). Since the Hubble constant is fixed by
the constraint equation,
\begin{equation}
 h = {(\omega_b + \omega_c)^{1/2} \over ( 1 - \Omega_k - \Omega_\Lambda)^{1/2}},  \label{2.2}
\end{equation}
it is almost unconstrained by the CMB data. 

The lower two panels in Fig. 3a show the constraints on the parameter
combinations $w_c$--$\omega_b$ and $n_s$--$\omega_b$. These 
panels illustrate the tensor degeneracy: including a
tensor component significantly broadens the allowed ranges
of parameters. For example, values of $\omega_b$ that are more than 
twice the value favoured from primordial nucleosynthesis are
allowed by the CMB data (Efstathiou 2001).

Fig. 3b shows likelihood contours using the CMB data alone, but
computed with the natural variables of the galaxy power spectrum
analysis as in Fig. 1. The parameter combination $\Omega_m h$ that
essentially fixes the shape of the matter power spectrum is
extremely unnatural for an analysis of the 
CMB anisotropies. Since $\Omega_m h \equiv (\omega_b +
\omega_c)/h$, the indeterminacy in $h$ arising from the geometrical
degeneracy smears the likelihoods along the direction of
$\Omega_m h$. The wide range of allowed values of $\omega_b/\omega_m$
and the tight correlation with $n_s$ is a consequence of the tensor
degeneracy.

\subsection{Combining the CMB and 2dFGRS likelihoods}

\begin{table*}
\bigskip

\centerline{\bf \ \ \ \ \ \ \ \ \ \ Table 1:  Parameters values and errors}

\begin{center}
\begin{scriptsize}

\begin{tabular}{lccccccc} \hline \hline
\smallskip 
  & & & & & \multicolumn{3}{c}{approximate $\pm 2\sigma$ parameter ranges} \cr
  & Fit A & Fit B & Fit C& Fit D &   Fit A & Fit C & Fit D \cr
  & CMB alone & CMB+2dFGRS & CMB+2dFGRS & CMB+2dFGRS+BBN& CMB alone & 
CMB+2dFGRS & CMB+2dFGRS+BBN \cr
  & + tensor & no tensor & + tensor & + tensor & + tensor & + tensor & + tensor \cr 
 & & & & & & &  \cr
 $\omega_b$ & $0.020$ & $0.021$ & $0.027$ & $0.020$ & $0.016$--$0.045$ &
$0.018$--$0.034$ & $0.018$--$0.022$  \cr 
 $\omega_c$ & $0.13$ & $0.12$ & $0.085$  & $0.10$ & $0.03$--$0.18$& 
$0.07$--$0.13$ & $0.08$--$0.13$ \cr 
 $n_s$ & $0.96$ & $1.00$ & $1.20$ & $1.04$  & $0.89$--$1.49$& 
$0.95$--$1.31$& $0.95$--$1.16$\cr 
 $\Omega_k$ & $-0.04$ & $0.001$ & $-0.030$ & $-0.013$& $-0.68$--$0.06$&
$-0.05$--$0.04$ & $-0.05$--$0.04$  \cr 
 $\Omega_\Lambda$ & $0.43$ & $0.71$ & $0.80$ & $0.73$ & $<0.88$& $0.65$--$0.85$ & $0.65$--$0.80$  \cr 
 $\tau_{opt}$ & 0 & 0 & 0 & 0 & $<0.5$& $<0.5$ & $<0.5$    \cr 
 $n_t$ &  &  - & $-0.10$ & $0.13$& & &     \cr 
 $\overline r$ & 0 & -& $0.60$& $0.20$& $<0.98$ & $<0.87$ & $<0.82$ \cr
 $r_{10}$ & 0 & - & $1.24$ & $0.26$ & & & \cr
 $\omega_b/\omega_m$ & $0.14$ & $0.15$ & $0.24$& $0.17$& $0.10$-$0.40$&
$0.13$--$0.28$ & $0.13$--$0.22$ \cr
 $\Omega_m h$ & & $0.21$ & $0.16$ & $0.19$ &  & $0.12$--$0.22$ & $0.16$--$0.21$\cr
 $h$ &  & $0.69$ & $0.71$& $0.66$& & $0.60$--$0.86$ & $0.61$--$0.84$ \cr
\hline
\end{tabular}
\end{scriptsize}
\end{center}

\end{table*}

Fig. 3b is interesting because it shows that the CMB likelihoods in
three of these plots are complementary to those of the 2dFGRS analysis
($\omega_b/\omega_m$--$\Omega_mh$, $n_s$ --$\Omega_m h$,
$\omega_b/\omega_m$ -- $n_s$). The addition of the 2dFGRS constraints
breaks both the geometrical and tensor degeneracies, resulting in
strong constraints on $\omega_b$, $\omega_c$, $\Omega_\Lambda$ and
$h$. The way that this works is evident from Figs 1 and 3b: the
constraints on $n_s$ from the 2dFGRS  help to break the tensor
degeneracy by excluding high values of $\omega_b$ and low values of
$\omega_c$. The resulting values of $\omega_b$ and $\omega_c$ fix the
Hubble radius at the time that matter and radiation have equal
density, which in turn largely fixes the shape of the CDM transfer
function in physical Mpc.  Comparing with the power spectrum of the
2dFGRS in $h^{-1}{\rm Mpc}$ constrains the Hubble constant, thus
breaking the geometrical degeneracy.

The lower panels in Fig. 3 show the results of combining the CMB and
2dFGRS likelihoods. The results are striking, showing a significant
tightening of the constraints in each plot.  Table 1 lists parameters
corresponding to maximum likelihood fits to the data and the
approximate $\pm 2\sigma$ ranges of each parameter.  The second column lists
the maximum likelihood fit to the CMB alone (fit A). The parameters of this
fit are identical whether or not a tensor component is included.
 The third 
and fourth columns (fits B and C) list the maximum likelihood fits to the
CMB and 2dFGRS data excluding and including a tensor mode. The fifth column
(fit D) adds the constraint from big-bang nucleosynthesis (BBN)
of a Gaussian distribution for $\omega_b$ centred at $\omega_b=0.020$
with a dispersion of $\Delta \omega_b = 0.001$ (Burles \etal 2001).

The parameters of fit B,
which provides a perfectly acceptable fit to the data,
are very close to those of the standard `concordance' cosmology
({\it e.g.} Bahcall \etal 1999). In particular, the baryon density 
is compatible with the primordial nucleosynthesis value, and the
Hubble and cosmological constants are compatible with 
more direct observational estimates. The CMB power spectrum for this 
solution is plotted as the solid line in Fig. 2 and the linear matter
power spectrum is plotted together with the 2dFGRS data points in Fig. 4.
Both curves provide acceptable fits to the data. Fit B has a
low baryon fraction of  $\omega_b/\omega_m = 0.15$. As a consequence, the
amplitudes of the baryonic features in the matter power spectrum are 
almost imperceptibly small (see Fig. 4).

\begin{figure}

\vskip 2.9 truein

\includegraphics{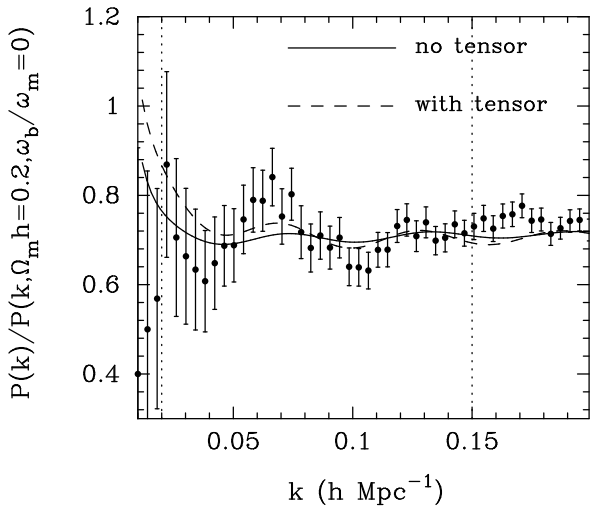}

\caption
{The points show the galaxy power spectrum of the 2dFGRS measured by
P01 divided by the power spectrum of a scale-invariant CDM model with
$\omega_b=0$, $\Omega_mh =0.2$. The error bars are computed from the
diagonal components of the covariance matrix. The lines show the
linear matter power spectra of the maximum likelihood fits to the
combined CMB and 2dFGRS power spectra after convolution with the
spherically averaged window function of the survey.  The solid line
shows fit B from Table 1 (no tensor component). The dashed line shows
fit C (including a tensor component).}
\label{figure4}
\end{figure}

Allowing a tensor component produces a slightly better fit to the
data, but the parameters are less concordant with other observations
(Fit C, Table 1). The CMB power spectrum for this model is plotted as
the dashed line in Fig. 2. According to this solution, a significant
part of the COBE anisotropies comes from a tensor component. The
baryon density of fit C is $\omega_b = 0.027$ and is well outside the
range of values inferred from primordial nucleosynthesis. The matter
power spectrum for this model is plotted as the dashed line in
Fig. 4. This shows clearly what is happening with this solution. The
apparent wiggles in the 2dFGRS power spectrum pull the solution
towards a high baryon fraction.  However, to produce a good fit to the
CMB anisotropies with a high baryon fraction, the tensor degeneracy of
Fig. 3 requires high values of $n_s$ and significant tensor
anisotropies. The likelihood ratio of fits B and C is ${\cal
L}_B/{\cal L}_C = 0.34$ and so fit C is only marginally preferred over
fit B. In two of the panels from Fig. 3c and d, the likelihood
distributions have two peaks centred at the parameters of fits B
and C. Adding the BBN constraint on $\omega_b$ (fit D) selects one
of these peaks with parameters close to those of fit B.

Fits B and C predict a lower normalization for the present day matter
power spectrum than implied by the local abundance of rich clusters of
galaxies. In a recent analysis of the number density distribution of
rich clusters as a function of X-ray temperature, Pierpaoli, Scott and
White (2001) deduce 
\begin{equation}
\sigma_8 = (0.495^{+0.034}_{-0.037})
\Omega_m^{-0.60}, \label{2.4}
\end{equation} 
where $\sigma_8$ is the rms fluctuation in the  mass  density
distribution averaged in spheres of radius $8 \hmpc$. 
Fit B gives
$\sigma_8 = 0.72$ and fit C gives $\sigma_8=0.61$, whereas equation
(\ref{2.4}) implies $\sigma_8=1.04$ and $\sigma_8=1.20$
respectively. Most of the error in equation (\ref{2.4}) comes from
uncertainties in the cluster mass-X-ray temperature relation
and it is not clear whether the quoted error reflects the true
uncertainties. A number of effects could boost the best fitting values
of $\sigma_8$, 
for example,  a  realistic value for $\tau_{opt}$ or
possible calibration errors in the CMB data might affect $\sigma_8$
at the $10$-$20\%$ level. Such effects may reconcile fit B with the
cluster data, but are probably not large enough to explain the
discrepancy with fit C. Furthermore, 
as we have discussed above, the discrepancy with
the primordial nucleosynthesis value of $\omega_b$ provides
another reason to disfavour fit C.

\section{Discussion}

The results of this paper are based on the key assumption that the
galaxy power spectrum on large scales (wavenumbers $k < 0.15 \hmpci$)
is proportional to the linear matter power spectrum. Under this
assumption, we have shown that the galaxy power spectrum of the 2dFGRS
can be used to partially break the two major parameter degeneracies
inherent in the analysis of CMB anisotropies.  The limits on the
scalar spectral index from the 2dFGRS help to break the tensor
degeneracy. The resulting constraints on the matter density provide a
measure of a standard physical distance (the Hubble radius at the time
that matter and radiation have equal density). This standard length
constrains the Hubble constant and so breaks the geometrical
degeneracy.

The resulting constraints are in remarkable agreement with the baryon
density inferred from primordial nucleosynthesis (Burles and Tytler
1998ab), estimates of the Hubble constant from the HST Hubble key
project (Freedman \etal 2001) and evidence for a non-zero cosmological
constant from obserations of distant Type Ia supernovae (Perlmutter
\etal 1999; Riess \etal 1999). The best fit model excluding a tensor
component has parameters that are very close to those of the standard
`concordance' cosmology (Bahcall \etal 1999).  However, the combined
CMB+2dFGRS data provide weak upper limits on a tensor component (Table 1)
and other solutions are possible which have a higher baryon density
and baryon fraction. These solutions conflict with the limits on
$\omega_b$ from primordial nucleosynthesis and require a scalar
spectral index $n_s > 1$. The model with high $n_s$ and high $\omega_b$
provides a somewhat closer match to the apparent  `wiggles' in
the galaxy power spectrum at wavenumbers $k \sim 0.08 \hmpci$ and $k
\sim 0.12 \hmpci$ than is achieved by the scalar only model
(Fig. 4). Neither model fully matches the data points, however, and it is
plausible that the apparent features are enhanced by the noise. New 
power spectrum data from the 2dFGRS and the SDSS will soon allow
us to test  this hypothesis.
It is particularly encouraging that the
combination of the 2dFGRS and CMB data  yields strong evidence for a
cosmological constant in the range $0.65 \simlt \Omega_\Lambda \simlt
0.85$ based on completely different arguments to those applied to distant
Type Ia supernovae. This significantly strengthens the case in favour of
an accelerating universe.

\vskip 0.15 truein

\noindent
{\bf Aknowledgments:} GE thanks Caltech for the award of a Moore
Scholarship. The CMB power spectra in this paper were computed using
the CMBFAST code of Seljak \& Zaldarriaga (1996).

\end{document}